\documentclass{aa520}
\usepackage{graphicx}
\usepackage{txfonts}
\usepackage{natbib}
\bibpunct{(}{)}{;}{a}{}{,}
\begin{document}

\title{Quasars near the line of sight towards Q~0302$-$003 and the transverse proximity effect
  \thanks{Based on observations collected
    at the European Southern Observatory, Chile (Proposals 070.A-0425 and
    074.A-0273). Data collected under Proposal 068.A-0194 was obtained from
    the ESO Science Archive. Based on observations made with the NASA/ESA
    Hubble Space Telescope, obtained from the data archive at the Space
    Telescope Institute. STScI is operated by the association of Universities
    for Research in Astronomy, Inc. under the NASA contract NAS 5-26555.}
} 

\titlerunning{Quasars near Q~0302$-$003 and the transverse proximity effect}
\author{G.~Worseck \and L.~Wisotzki}
\institute{Astrophysikalisches Institut Potsdam, An der Sternwarte 16, 14482 Potsdam, Germany}
\offprints{G.~Worseck, \email{gworseck@aip.de}}
\date{Received... / Accepted...}

\abstract{We report the discovery of the faint ($V\simeq 21.7$) quasar
  QSO~03027$-$0010 at $z=2.808$ in the vicinity of Q~0302$-$003,
  one of the few quasars observed with STIS to study intergalactic \ion{He}{ii}
  absorption. Together with another newly discovered QSO at $z=2.29$, there
  are now 6 QSOs known near the line of sight towards Q~0302$-$003, of which
  4 are located within the redshift region $2.76 \la z \la 3.28$ covered by
  the STIS spectrum. 
  We correlated the opacity variations in the \ion{H}{i} and
  \ion{He}{ii} Lyman forest spectra with the locations of known quasars.
  There is no significant proximity effect in the \ion{H}{i} Ly$\alpha$ forest 
  for any of the QSOs, except for the well-known line of sight effect for 
  Q~0302$-$003 itself.
  By comparing the absorption properties in \ion{H}{i} and \ion{He}{ii}, we 
  estimated the fluctuating hardness of the extragalactic UV radiation field 
  along this line of sight. We find that close to each foreground quasar,
  the ionizing background is considerably harder than on average. In
  particular, our newly discovered QSO~03027$-$0010 shows such a hardness
  increase despite being associated with an overdensity in the \ion{H}{i} 
  Lyman forest.
  We argue that the spectral hardness is a sensitive physical measure 
  to reveal the influence of QSOs onto the UV background even over scales
  of several Mpc, and that it breaks the density degeneracy hampering the
  traditional transverse proximity effect analysis. 
  We infer from our sample that there is no need for significantly anisotropic
  UV radiation from the QSOs.
  From the transverse proximity effect detected in the sample 
  we obtain minimum quasar lifetimes in the range $\sim$10--30~Myr.

} 

\maketitle

\section{Introduction}

Observations of high-redshift quasars enable us to study the intergalactic
medium (IGM) along their lines of sight via the absorption of quasar radiation
by various chemical elements in different ionization stages. Hydrogen and
helium are by far the most abundant elements in the universe and the
Ly$\alpha$ transitions of \ion{H}{i} and \ion{He}{ii} in an incompletely ionized
medium cause a Gunn-Peterson trough at redshifts smaller than the emission
redshift of the observed quasar \citep{gunn65}. If the IGM is highly ionized,
a plethora of discrete absorption lines stemming from the remaining neutral
fraction is visible both in \ion{H}{i} Ly$\alpha$ and \ion{He}{ii} Ly$\alpha$,
giving rise to the name Ly$\alpha$ forest.

The intensity of the metagalactic UV radiation field at a characteristic
frequency (typically the ionization energy of a given element) varies in time
due to the temporal evolution of the source population. Spatial fluctuations
induced by the discreteness of the source population have been examined for
randomly distributed sources and absorbers by \citet{zuo92} and
\citet{fardal93}. They are also treated in recent numerical simulations trying
to quantify their impact on the Ly$\alpha$ flux power spectrum
\citep{croft99,meiksin04,croft04,mcdonald05}. The spatial fluctuations of the
UV radiation field computed from these simulations along random lines of sight
are generally gentle (few per cent around the mean) and occur on large scales
($\ga 100$~Mpc comoving) at $z<4$.  This simple picture with only mild
large-scale fluctuations is expected to change considerably for lines of sight
that pass close to sources of the UV background, such as luminous quasars
\citep{fardal93,croft04,mcdonald05}. The source flux acts as a local
enhancement of the UV radiation field. As a consequence, the absorbers in the
region affected by this excess flux will be statistically more ionized than
the rest of the Ly$\alpha$ forest along the line of sight, resulting in a
statistically increased transmission, a radiation-induced `void' in the
\ion{H}{i} Ly$\alpha$ forest in the vicinity of the UV source. This so-called
proximity effect has been detected with high statistical significance in lines
of sight towards luminous quasars \citep[e.g.][]{bajtlik88,scott00}. However,
a transverse proximity effect created by foreground ionizing sources nearby
the line of sight has not been clearly detected in the \ion{H}{i} Ly$\alpha$
forest. The full range of possible results extends from large voids claimed to
be due to the transverse proximity effect by 
\citet[][however see \citealt{dobrzycki91b}]{dobrzycki91} and \citet{srianand97}, 
over marginal detections \citep{fernandez-soto95,liske01} to non-detections
\citep{crotts89,moller92,crotts98,schirber04}. \citet{croft04} measured the
average transverse Ly$\alpha$ transmission from all projected quasar pairs in
the SDSS DR1 and even found excess absorption near foreground quasars instead
of the expected excess transmission caused by the transverse proximity effect.

The detectability of the transverse proximity effect is hampered by several
systematic effects. Anisotropic radiation of quasars has been invoked to
explain redshift offsets between the void and the foreground quasar
\citep{dobrzycki91} or to explain the lack of the transverse proximity effect
\citep{crotts89,moller92,schirber04}. Also quasar variability affects the
detection of the proximity effect \citep{schirber04}. Finally, the possible
gravitational clustering around quasars that are assumed to reside in the
densest environments may dilute the proximity effect
\citep{loeb95,schirber04}. According to \citet{schirber04} only a combination
of these systematic effects may explain the apparent absence of the transverse
proximity effect.

\ion{He}{ii} Ly$\alpha$ $303.78$~\AA~ absorption can be studied only towards a
few lines of sight to date because most of the quasars at $z>2$ have
intervening optically thick Lyman limit systems that truncate the flux in the
observable \ion{He}{ii} Ly$\alpha$ wavelength range in the far UV. The
observations of the lines of sight towards Q~0302$-$003 at $z=3.285$
\citep{jakobsen94,hogan97,heap00}, \object{PKS~1935$-$692} at $z=3.18$
\citep{anderson99} and recently \object{SDSS~J2346$-$0016} at $z=3.50$
\citep{zheng04b} show in most parts of their \ion{He}{ii} absorption spectra
very strong absorption at $z>3$ that is consistent with a Gunn-Peterson trough
($\tau_\ion{He}{ii}>3$). In contrast, the three lines of sight at $z<3$ probed
so far towards \object{HS~1700$+$6416} at $z=2.72$
\citep{davidsen96,reimers04}, \object{HE~2347$-$4342} at $z=2.885$
\citep{reimers97,kriss01,smette02,shull04,zheng04} and recently
\object{QSO~1157$+$3143} at $z\simeq 3$ \citep{reimers05} display patchy
intergalactic \ion{He}{ii} absorption with voids ($\tau_\ion{He}{ii}\lesssim
1$) and troughs ($\tau_\ion{He}{ii}>3$) that evolves to a \ion{He}{ii}
Ly$\alpha$ forest at $z<2.7$ resolved with FUSE
\citep{kriss01,shull04,zheng04,reimers04}. In combination with the observed 
evolution of \ion{H}{i} line widths \citep{schaye00,ricotti00,theuns02a} 
these observations point to a late \ion{He}{ii} reionization between 
$z\sim 2.7$ and $z\sim 3$.

By comparing the \ion{H}{i} absorption with the corresponding \ion{He}{ii}
absorption one can estimate the hardness of the ionizing radiation field that
penetrates the IGM, since \ion{H}{i} is ionized at $h\nu>13.6$~eV, whereas
\ion{He}{ii} is ionized at $h\nu>54.4$~eV. The amount of \ion{He}{ii} compared
to \ion{H}{i} gives a measure of the spectral hardness. Already low-resolution
\ion{He}{ii} observations obtained with HST indicated a fluctuating radiation
field in the voids (hard) and the troughs (soft)
\citep{reimers97,heap00,smette02}. The recent high-resolution FUSE
observations of the \ion{He}{ii} Ly$\alpha$ forest reveal large fluctuations
on very small scales of $\Delta z\sim 10^{-3}$
\citep{kriss01,shull04,reimers04}. Due to the hard ionizing field required in
the \ion{He}{ii} voids that is consistent with the integrated radiation of a
surrounding quasar population, these \ion{He}{ii} voids have been interpreted
as the onset of \ion{He}{ii} reionization in Str\"{o}mgren spheres around hard
\ion{He}{ii} photoionizing sources along or near the line of sight
\citep{reimers97,heap00,smette02}. A subsequent survey for putative quasars
that may cause the prominent \ion{He}{ii} void at $z=3.05$ towards
Q~0302$-$003 yielded the quasar \object{QSO~03020$-$0014} located $6\farcm5$
away on the sky that coincides with this \ion{He}{ii} void
\citep{jakobsen03}. Thus, the \ion{He}{ii} void in Q~0302$-$003 is the first 
clear case of a transverse proximity effect due to a luminous quasar.

In this paper we report on results from a slitless spectroscopic quasar survey
that resulted in the discovery of another foreground quasar in the vicinity of
Q~0302$-$003. The structure of the paper is as follows. 
Sect.~\ref{observations} describes the observation and the supplementary data
employed for the paper. In Sect.~\ref{intensity_h1} and \ref{intensity_he2} we
examine the evidence for opacity variations in the \ion{H}{i} and \ion{He}{ii}
Ly$\alpha$ forest regions caused by intensity fluctuations in the UV radiation
field towards Q~0302$-$003 at 1~ryd and 4~ryd, respectively.  In
Sect.~\ref{hardness} we consider the possibility to detect the transverse
proximity effect via spectral hardness diagnostics of the ionizing radiation.
We argue that this is actually the most sensitive method, and we demonstrate
that essentially each quasar near the line of sight is associated with a local
hardening of the radiation field. Finally, we use these observations to
estimate a lower limit to the quasar lifetime (Sect.~\ref{lifetime}).
We present our conclusions in Sect~\ref{conclusions}.
Throughout the paper we adopt a flat cosmological model with
$\Omega_\mathrm{m}=0.3$, $\Omega_\Lambda=0.7$ and
$H_0=70$~$\mathrm{km}\,\mathrm{s}^{-1}\,\mathrm{Mpc}^{-1}$.

\section{Observations \& data reduction} \label{observations}

\subsection{Search for quasar candidates near Q~0302$-$003}

\begin{figure*}
\centering 
\includegraphics[scale=0.93]{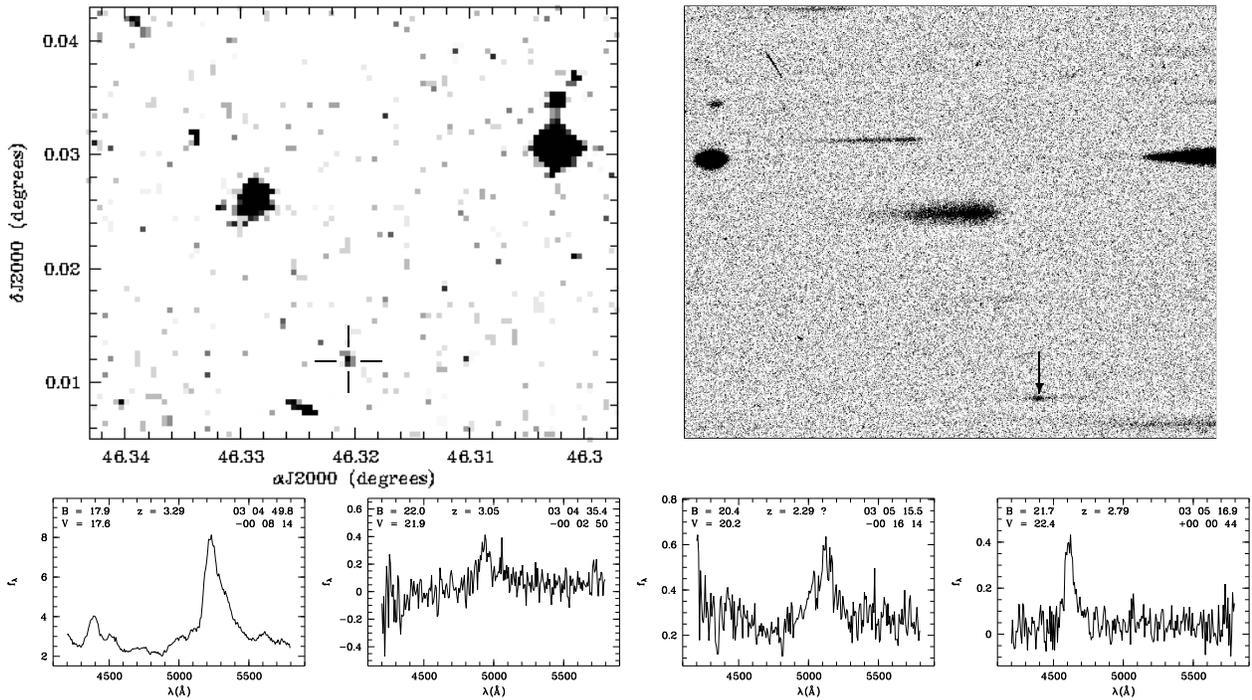}
\caption{\label{wfispec} Discovery of the two quasar candidates. The upper
left panel shows a $2.8\arcmin\times2.3\arcmin$ DSS1 image of the sky region
around the faint QSO~03027$-$0010 marked with a cross. The upper right panel
displays the corresponding region of the slitless WFI $B$ exposures ($0\degr$
rotation, 1800~s stacked observations). The emission line of QSO~03027$-$0010
is clearly visible (arrow). The lower panels show the combined and calibrated
slitless spectra of $z>2$ quasars found on the total $25\arcmin\times
33\arcmin$ field with their DSS1 position, redshift and magnitude estimates
(from left to right: Q~0302$-$003, QSO~03020$-$0014, QSO~03027$-$0027 and
QSO~03027$-$0010).}
\end{figure*}

In October 2002 and February 2003 we conducted a quasar survey with the ESO
Wide Field Imager (WFI) at the ESO/MPI $2.2$~m Telescope at La Silla in its
slitless spectroscopic mode \citep{wisotzki01}. The survey fields
of $25\arcmin\times 33\arcmin$ were centered on bright high-redshift QSOs, and
our aim was to find faint quasars in their vicinity. Details about this survey
will be given in a separate paper. Here we report the results of the field
centered on Q~0302$-$003 at $z=3.285$ observed on October 3 and 4, 2002, and we
only briefly summarize the survey technique. We recorded our slitless spectra
with the R50 grism ($\lambda/\Delta\lambda=30$--50 depending on the seeing) 
in the $B$ and the $V$ band by 3 dithered 600~s exposures
in 2 instrument rotations each, resulting in a total exposure time of 1~h per
band and yielding a limiting magnitude of $V_{\mathrm{lim}}\simeq 22$. The
filters were employed to reduce the sky background and the degree of crowding
by limiting the length of the recorded spectra. Crowding was further accounted for
by taking the spectra in two instrument rotations (0\degr~and 10\degr). The
combination of the R50 grism with the broadband $B$ and $V$ filters resulted
in a spectral coverage from the blue grism sensitivity cutoff at 4200~\AA\ to
5800~\AA, permitting us to search for Ly$\alpha$ emission in the redshift
range $2.46\lesssim z\lesssim 3.77$ and for \ion{C}{iv} emission in the range
$1.71\lesssim z\lesssim 2.74$.

All sources detected in the Digitized Sky Survey (which has approximately
the same depth as our slitless spectroscopic data)
were automatically extracted, yielding flux-calibrated low-resolution spectra of
$\sim 800$ objects in the field. The spectra were searched automatically for
emission-line objects. The lower panel of Fig.~\ref{wfispec} shows the
slitless spectra of four $z>2$ quasars that were found by our survey. One of them
was (unsurprisingly) Q~0302$-$003 itself, and we also unambiguously
rediscovered QSO~03020$-$0014 discovered recently by \citet{jakobsen03}. 
Our search revealed two further quasar candidates in the vicinity of
Q~0302$-$003 that fall into our redshift range of interest. 
The first showed clearly Ly$\alpha$ emission at $z\simeq 2.79$ 
whereas the second displayed an emission line which we tentatively identified
as \ion{C}{iv} at $z\simeq 2.29$.

\begin{figure*}
\sidecaption
\includegraphics[width=12cm]{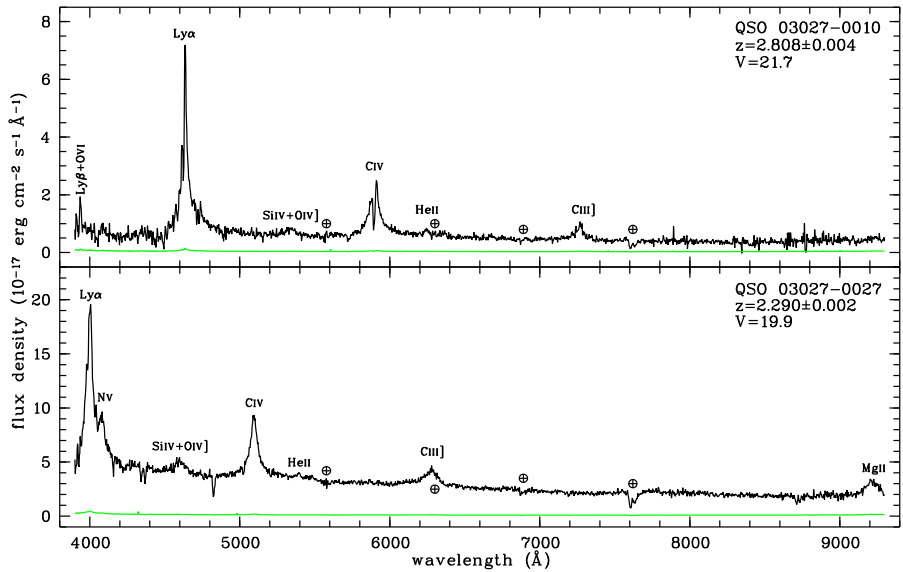}
\caption{\label{qsospec}VLT/FORS2 spectra of the two discovered QSOs. The
spectra are shown in black lines together with their corresponding $1\sigma$
noise arrays (green lines). Detected emission lines and major atmospheric
artifacts are indicated.}
\end{figure*}

\begin{table*}
\caption{Observing log for the spectroscopic follow-up.}
\label{observinglog}
\centering
\begin{tabular}{lllllllll}
\hline\hline\noalign{\smallskip}
Object &$\alpha$~(J2000) &$\delta$~(J2000) &Night &Grism &Slit&Exposure &Airmass&Seeing  \\
\noalign{\smallskip}\hline\noalign{\smallskip}
QSO~03027$-$0010&$03^\mathrm{h}05^\mathrm{m}16\fs80$
&$+00\degr00\arcmin45\farcs1$ & 17 Nov 2004 & 300V & $1\farcs0$ &600~s  &$1.10$ &0\farcs9 \\
QSO~03027$-$0027&$03^\mathrm{h}05^\mathrm{m}15\fs62$
&$-00\degr16\arcmin14\farcs4$ & 17 Nov 2004 & 300V & $1\farcs0$ &300~s  &$1.10$ &0\farcs7 \\
Q~0302$-$003    &$03^\mathrm{h}04^\mathrm{m}49\fs71$ 
&$-00\degr08\arcmin13\farcs0$ & 19 Nov 2004 & 600B & $1\farcs0$ &1800~s &$1.18$ &$\lesssim1\farcs9$\\
\noalign{\smallskip}\hline
\end{tabular}
\end{table*}

\subsection{Spectroscopic follow-up}

Follow-up spectroscopy of these two quasar candidates was obtained with the
Focal Reducer/Low Dispersion Spectrograph 2 (FORS2) on ESO VLT UT1/Antu in
Visitor Mode on November 17, 2004. The sky was clear throughout the night. The
spectra were taken with the 300V grism and a 1\arcsec~slit kept at the
parallactic angle, resulting in a spectral resolution of $\sim10$~\AA~FWHM. In
order to maximize throughput, the order separation filter was omitted, leading
to possible order overlap at $\lambda>6600$~\AA. Exposure times were adjusted
to yield a $S/N$ of $\sim20$ in the quasar continuum. We also recorded a FORS2
spectrum of Q~0302$-$003 with the 600B grism on November 19, 2004, at a
resolution of $\sim4.5$~\AA~FWHM. Seeing conditions were poor and highly
variable during the observations of Q~0302$-$003, resulting in some slit losses 
but still a $S/N$ of $\sim 70$ in the Ly$\alpha$ forest could be
achieved. The spectra were calibrated in wavelength against the FORS2
He/Ne/Ar/HgCd arc lamps and spectrophotometrically calibrated against the HST
standard stars Feige~110 and GD~108. Data reduction was performed with
standard IRAF tasks. The spectra were optimally extracted using the algorithm
introduced by \citet{horne86}. Table~\ref{observinglog} summarizes our
spectroscopic follow-up observations.

Figure~\ref{qsospec} shows the calibrated spectra of the two candidates. Both
are confirmed to be quasars at redshifts correctly estimated in
the slitless spectra. Since the quality of our FORS2 aquisition images is
superior to the DSS image, we refined the QSO positions (see
Table~\ref{observinglog}) and denote these as QSO~03027$-$0010 and
QSO~03027$-$0027 in the following, according to IAU conventions.

There is a strong associated absorption line system in the spectrum of
QSO~03027$-$0010 at $z_\mathrm{abs}=2.804\pm 0.001$ visible in Ly$\alpha$,
\ion{N}{v} and \ion{C}{iv}. A \ion{C}{iv} system is present at
$z_\mathrm{abs}=2.696$ that is also visible in \ion{Si}{iv} and that
corresponds to a strong Ly$\alpha$ absorption line in the Ly$\alpha$
forest. Moreover, a \ion{Mg}{ii} absorber may be present at
$z_\mathrm{abs}=0.750$.

QSO~03027$-$0027 shows a strong metal line system at $z_\mathrm{abs}=2.115$,
easily detectable in \ion{Mg}{ii}, \ion{C}{iv}, \ion{Si}{iv} and
\ion{C}{ii}. Another weaker \ion{C}{iv} system may be present at
$z_\mathrm{abs}=2.243$.

Table~\ref{redshifts} shows our quasar redshift determination based on all
detectable emission lines. The $S/N$ of both spectra generally prevents a
clear detection of low-ionization lines, such as \ion{O}{i}$+$\ion{Si}{ii} or
\ion{O}{iii}]. The redshift measurement based on the \ion{Mg}{ii} line in
QSO~03027$-$0027 is inaccurate due to the decreasing resolving power of the
grism at the longest recordable wavelengths. Consequently, the redshifts have
to be measured in high-ionization lines which suffer from systematic
blueshifts \citep{gaskell82,tytler92,mcintosh99}. The measurements in the
spectrum of QSO~03027$-$0010 are further affected by the associated absorption
system. Having these caveats in mind, we adopt a redshift of $z=2.290\pm
0.002$ for QSO~03027$-$0027 and a redshift of $z=2.808\pm 0.004$ for
QSO~03027$-$0010.

Apparent magnitudes of both quasars were derived based on the available
spectrophotometry, yielding $V=21.7$ for QSO~03027$-$0010 and $V=19.9$ for
QSO~03027$-$0027.

\subsection{Continuum and redshift of Q~0302$-$003}

\begin{table}
\caption{Redshift determination of the two newly discovered QSOs.}
\label{redshifts}
\scriptsize \centering
\begin{tabular}{lcccc}
\hline\hline\noalign{\smallskip} 
Emission &\multicolumn{2}{c}{QSO~03027$-$0010} & \multicolumn{2}{c}{QSO~03027$-$0027} \\
line &$\lambda$~[\AA] &$z$ &$\lambda$~[\AA] &$z$ \\
\noalign{\smallskip} \hline \noalign{\smallskip} 
Ly$\beta$+\ion{O}{vi}&$3936\pm 5$ &$2.810\pm 0.005$ &-- &-- \\
Ly$\alpha$    &$4627\pm 7$  &$2.806\pm 0.006$ &$4004\pm 3$ &$2.294\pm 0.003$\\
\ion{N}{v}    &--           &-- &$4081\pm 3$ &$2.291\pm 0.003$\\
\ion{Si}{iv}+\ion{O}{iv}]&$5335\pm 10$ &$2.812\pm 0.007$ &$4598\pm 8$ &$2.284\pm 0.006$\\ 
\ion{C}{iv}   &$5896\pm 11$ &$2.806\pm 0.007$ &$5095\pm 2$ &$2.289\pm 0.001$\\ 
\ion{He}{ii}  &$6244\pm 3$  &$2.806\pm 0.002$ &$5393\pm 3$ &$2.288\pm 0.002$\\ 
\ion{C}{iii}] &$7268\pm 4$  &$2.808\pm 0.002$ &$6278\pm 5$ &$2.289\pm 0.003$\\ 
\ion{Mg}{ii}  &-- &-- &$9223\pm 15$ &$2.296\pm 0.005$\\
\noalign{\smallskip} \hline
\end{tabular}
\end{table}

\begin{table*}
\caption{Known $z>2$ foreground quasars within 30\arcmin~of
Q~0302$-$003 ($z=3.285$). We list positions, redshifts, $V$
magnitudes, projected angular distances $\vartheta$ from Q~0302$-$003, and the
corresponding transverse proper distances $d_\perp(z)$ at
the emission epoch of each foreground quasar.}
\label{foregroundquasars}
\centering
\begin{tabular}{lllllllll}
\hline\hline \noalign{\smallskip} QSO & Abbr. & $\alpha$~(J2000) &$\delta$~(J2000) &$z$ &$V$
&$\vartheta$~[\arcmin] &$d_\perp$~[Mpc]&discovery paper\\ 
\noalign{\smallskip} \hline \noalign{\smallskip} 
QSO~03022$-$0023 & F & $03^\mathrm{h}04^\mathrm{m}45\fs94$
             &$-00\degr11\arcmin38\farcs2$ 
             &$2.142$&$22.5$ &$3.55$ &$1.77$ &\citet{jakobsen03}\\ 
QSO~03027$-$0027 & E & $03^\mathrm{h}05^\mathrm{m}15\fs62$
             &$-00\degr16\arcmin14\farcs4$ 
             &$2.290$&$19.9$ &$10.31$ &$5.08$ &this paper\\
QSO~03027$-$0010 & B & $03^\mathrm{h}05^\mathrm{m}16\fs80$
             &$+00\degr00\arcmin45\farcs1$ 
             &$2.808$&$21.7$ &$11.24$ &$5.29$ &this paper\\
Q~0302-D113  & D & $03^\mathrm{h}04^\mathrm{m}30\fs33$ 
             &$-00\degr08\arcmin11\farcs4$
             &$2.920$&$24.3$ &$4.85$ &$2.26$ &\citet{steidel03}\\
QSO~03020$-$0014 & A & $03^\mathrm{h}04^\mathrm{m}35\fs37$
             &$-00\degr02\arcmin50\farcs9$ 
             &$3.050$&$20.5$ &$6.46$ &$2.97$ &\citet{jakobsen03}\\ 
Q~0301$-$005 & C & $03^\mathrm{h}03^\mathrm{m}41\fs05$ 
             &$-00\degr23\arcmin21\farcs8$ 
             &$3.231$&$17.8$ &$22.89$ &$10.34$ &\citet{barbieri86}\\ 
\noalign{\smallskip} \hline \noalign{\smallskip} 
\end{tabular}
\end{table*}

In order to be able to use the FORS2 spectrum of Q~0302$-$003 for measuring 
its transmission properties in the \ion{H}{i} Ly$\alpha$ forest region,
we had to obtain a working estimate of the continuum. 
This was established by fitting a cubic spline to selected high-transmission
regions of the data, using the DIPSO software under Starlink
\footnote{The authors acknowledge the data analysis facilities provided by the 
Starlink Project which is run by CCLRC on behalf of PPARC.}. 
Due to the low resolution of the spectrum, the absorption lines are heavily 
blended. We checked our continuum fit for systematic errors using 
line lists from high-resolution spectra 
published by \citet{hu95} from Keck HIRES data and by \citet{kim02} from 
VLT UVES data. The list from \citet{hu95}
covers the redshift range $2.627\le z\le 3.110$, whereas the list from
\citet{kim02} extends between $2.957\le z\le 3.235$. The two line lists were
used to create artificial Ly$\alpha$ forest spectra which were subsequently
convolved with the line spread function of the FORS2 600B grism and rebinned to
$1.5$~\AA/pixel. The result was an excellent correspondence (r.m.s. error $\sim 3$\%) 
between our data and the simulated spectra, indicating that our fit did not grossly
underestimate the true continuum. The only regions where our continuum estimate
is significantly too low by $\sim 8$\% are the wavelength range largely devoid of
Ly$\alpha$ absorption from $\sim 5050$~\AA~to $\sim 5100$~\AA, known as the
Dobrzycki \& Bechtold void \citep{dobrzycki91} and a small region at $\sim
4480$~\AA\ already in the line wing of the Ly$\beta+$\ion{O}{vi} line of
Q~0302$-$003 where continuum fitting becomes complicated.

We also checked the line lists of \citet{hu95} and \citet{kim02} for metal
line contamination of the \ion{H}{i} Ly$\alpha$ forest. Only a few identified
systems contaminate the Ly$\alpha$ forest. These are narrow lines with low
column densities, which are heavily blended with \ion{H}{i} lines at our low
resolution. The absorption by metals is therefore negligible and we will treat
the whole absorption in the spectral range of the \ion{H}{i} Ly$\alpha$ forest
as genuine \ion{H}{i} Ly$\alpha$ absorption.

Our FORS2 spectrum of Q~0302$-$003 covers the wavelength range from the UV
cutoff to $\sim$6350~\AA~and has a $S/N$ of $\sim 75$ in the quasar continuum,
so we can measure the positions of the lines Ly$\beta+$\ion{O}{vi},
Ly$\alpha$, \ion{N}{v}, \ion{O}{i}$+$\ion{Si}{ii}, \ion{C}{ii} and
\ion{Si}{iv}+\ion{O}{iv}]. We base our redshift measurement on the
low-ionization lines \ion{O}{i}$+$\ion{Si}{ii} and \ion{C}{ii}, which should
yield the best possible estimate of the systemic redshift. We get a redshift
of $z=3.285\pm 0.001$ for the \ion{O}{i}$+$\ion{Si}{ii} line and a redshift of
$z=3.284\pm 0.001$ for the \ion{C}{ii} line, which is consistent with the
value of $3.286$ originally obtained by \citet{sargent89}. But we note that
their value was based on the Ly$\alpha$ and the \ion{C}{iv} emission line.

The sky region around Q~0302$-$003 was targeted by SDSS, and Q~0302$-$003 was
observed spectroscopically. We checked the SDSS DR3 database for the
spectroscopic redshift of Q~0302$-$003 and found the redshift to be too high,
$z_\mathrm{SDSS}\simeq 3.295\pm 0.001$. Given this large offset of $\Delta
z=0.009$, we obtained the SDSS spectrum of Q~0302$-$003 from the SDSS data
archive and recovered the redshift values for the low-ionization lines
\ion{O}{i}$+$\ion{Si}{ii} ($z=3.286\pm 0.002$) and \ion{C}{ii} ($z=3.284\pm
0.002$). We conclude that the SDSS redshift assignment for
Q~0302$-$003 is incorrect and adopt a redshift of $z=3.285$ for this quasar in
the following.

\subsection{HST/STIS data}

Q~0302$-$003 was one of the few high-redshift quasars observed successfully 
in the \ion{He}{ii} Ly$\alpha$ forest below $303.78$~\AA\ rest frame wavelength
with HST and its \emph{Space Telescope Imaging Spectrograph} at a resolution of $1.8$~\AA\ 
\citep[][hereafter refered to as H00]{heap00}. 
We retrieved these data from the HST archive and re-reduced them using 
CALSTIS v$2.13$ distributed with IRAF. As in the reduction by H00, 
we adjusted the background subtraction windows in order to correct for the
spatial variability of the STIS MAMA1 detector background. 
We adopt their flux normalization with a power law $f_{\nu}\propto\nu^{-\alpha}$ 
with $\alpha=2$, yielding $f_{\lambda}=\mathrm{const.}\simeq 2.1\times
10^{-16}\,\mathrm{erg}\,\mathrm{cm}^{-2}\,\mathrm{s}^{-1}\,\mathrm{Hz}^{-1}$. The
spectral index of 2 is consistent with the composite QSO EUV spectral index of
$\sim 1.8$ by \citet{telfer02}.

\subsection{Quasars in the vicinity of Q~0302$-$003} \label{systemicredshifts}

\begin{figure}
\centering \includegraphics{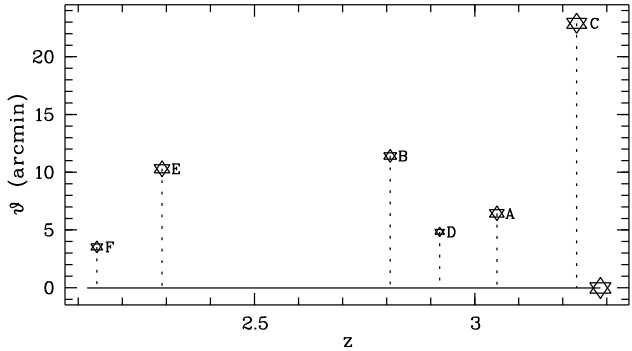}
\caption{\label{qsofield}Distribution of the foreground quasars from 
Table~\ref{foregroundquasars} with respect to Q~0302$-$003. Symbol size
indicates apparent optical magnitude.}
\end{figure}

Table~\ref{foregroundquasars} tabulates all 6 known quasars at $z>2$ within a
radius $<30$\arcmin~around Q~0302$-$003. Their distribution in redshift and separation 
from the central line of sight is shown in Fig.~\ref{qsofield}.
In addition to the QSOs discovered or rediscovered in our survey, we found 
two further quasars listed by \citet{veron03}. Spectra for these QSOs were 
retrieved from public archives, in order to measure the emission redshifts as 
consistently as possible.
For the sake of clarity, we will use a simplified nomenclature in the following 
and denote the foreground QSOs by the letters A \dots F as 
indicated in Table~\ref{foregroundquasars}.

Q~0301$-$005 (QSO~C) is a bright QSO known since long. It was observed by SDSS
and we obtained its spectrum from the SDSS data archive. Previous redshift
determinations were based on high-ionization lines \citep{sargent89}, so its
redshift of $z=3.223$ measured by \citet{sargent89} may have been
underestimated. The SDSS spectrum of this QSO has sufficient $S/N$ to measure
the low-ionization lines \ion{O}{i}$+$\ion{Si}{ii} and \ion{C}{ii}, yielding
a redshift of $z=3.231$ for this object in agreement with the SDSS redshift
assignment. Note the large shift of $\Delta z=0.008$ with respect to the
measurement based on high-ionization lines.

The two quasars discovered by \citet[][hereafter refered to as
J03]{jakobsen03} have spectra taken with FORS1 in the ESO/VLT Science
Archive. We performed an independent reduction and give here a refined
redshift (redshift uncertainty $0.002$) and the magnitude for
\object{QSO~03022$-$0023} (QSO~F). With $V = 22.5$ it is too faint to be detected
in our slitless spectroscopic survey material. Our estimate of redshift and
redshift error for QSO~03020$-$0014 (QSO~A) agrees with the values given in
J03.

\object{Q~0302-D113} (QSO~D) was published by \citet{steidel03} after the
compilation of the \citet{veron03} catalog. No spectrum is available for this
very faint ($R\simeq 24.6$) object. Its $V$ magnitude in
Table~\ref{foregroundquasars} is estimated from $R$.

\begin{figure*}
\centering \includegraphics[scale=0.95]{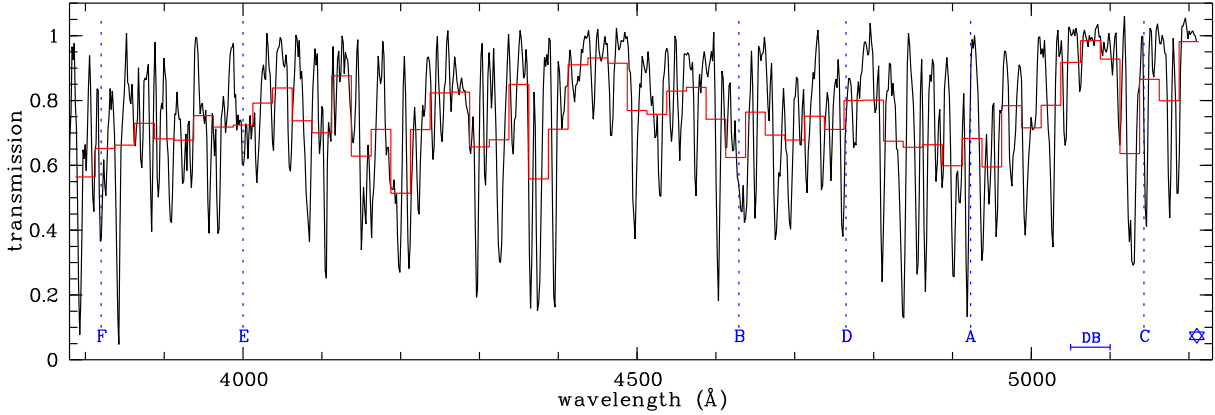}
\caption{\label{q0302forest} 
Transmission spectrum of the \ion{H}{i} Ly$\alpha$ \& Ly$\beta$ forest of
Q~0302$-$003. The spectrum is shown in black from 3780~\AA\ to 5210~\AA\
(corresponding to the Ly$\alpha$ emission peak of Q~0302$-$003). Blue letters and vertical 
dotted lines mark the foreground quasars from Table~\ref{foregroundquasars} projected into the forest. 
We also indicated the background quasar Q~0302$-$003 by a star symbol and the void identified by 
\citet{dobrzycki91} (DB). The binned red line is the average transmission in 25~\AA\ bins. 
\emph{[See the online edition of the Journal for a colour version of this figure.]}}
\end{figure*}

\section{No visible \ion{H}{i} transverse proximity effect} \label{intensity_h1}\label{trprox_h1}

We first conducted a simple visual search for locally enhanced transmission of
the \ion{H}{i} Lyman forest at the location of the foreground quasars. Figure
\ref{q0302forest} displays our low-resolution transmission spectrum of
Q~0302$-$003, together with the effective locations of the foreground quasars
projected into the forest. The proximity effect manifests itself as a
statistical increase in Ly$\alpha$ forest transmission near a quasar. Already
a quick examination of Fig.~\ref{q0302forest} shows that there are at least no
obvious \ion{H}{i} voids located at any of the foreground quasars that could
be revealed by visual inspection, with the possible exception of QSO~A which
is located very close to a local transmission maximum. There is no such
correlation for any of the other QSOs. There is, however, a significant increase
of Ly$\alpha$ forest transmission towards the Ly$\alpha$ line of the central QSO 
Q~0302$-$003 itself.

\begin{figure*}
\centering
\includegraphics[scale=0.95]{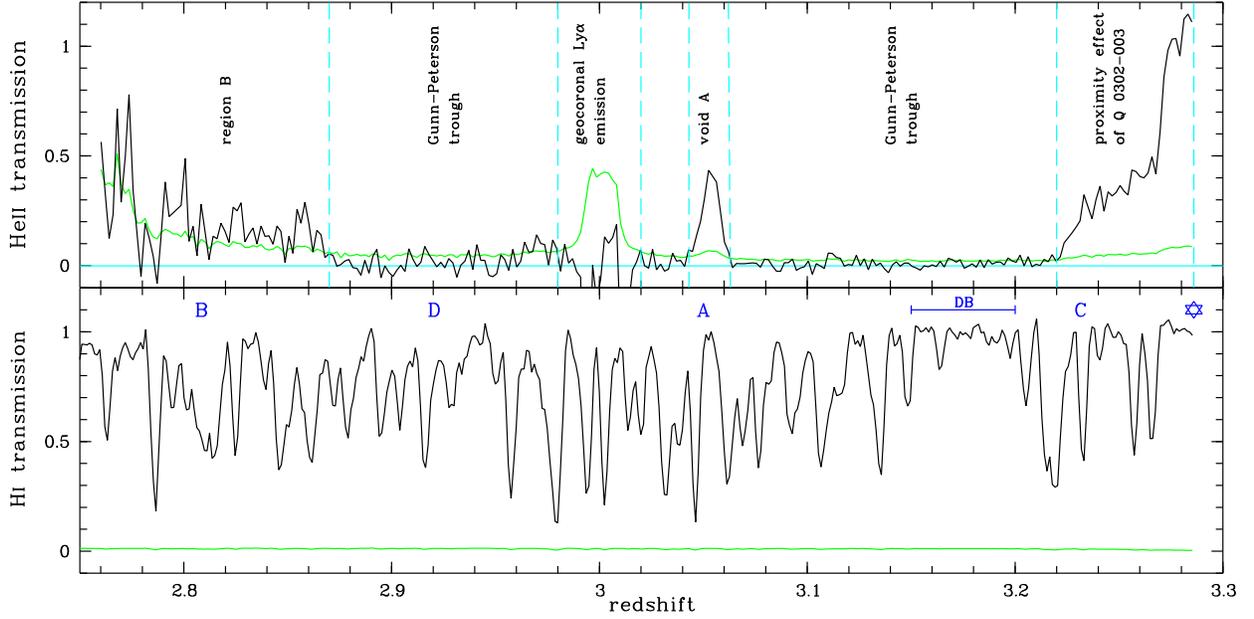}
\caption{\label{he2h1trans}
Comparison of the \ion{He}{ii} and the \ion{H}{i} Ly$\alpha$ absorption
spectrum towards Q~0302$-$003. Upper panel: Transmission in \ion{He}{ii} 
Ly$\alpha$ vs.\ redshift from the HST/STIS data obtained by H00. 
The green curve shows the $1\sigma$ uncertainty per pixel. The horizontal 
cyan line denotes the zero flux level. Vertical dashed lines divide the 
spectrum into several labeled regions of interest (see text). 
The lower panel displays the corresponding section of our FORS2 \ion{H}{i} 
Ly$\alpha$ transmission spectrum. Quasar and void locations are marked as in Fig.~\ref{q0302forest}. 
\emph{[See the online edition of the Journal for a colour version of this figure.]}}
\end{figure*}

Considering the luminosities of the foreground QSOs and their projected angular 
distances from the central line of sight, this lack of a transverse proximity
effect in \ion{H}{i} is no real surprise. 
We computed the parameter

\begin{equation}
\omega\left(z\right)=\sum_{j=1}^n\frac{f_{\nu_\mathrm{LL},j}}{4\pi J_\nu\left(z\right)}
      \frac{\left(1+z_j^\prime\right)^{-\alpha_j+1}}{\left(1+z_j\right)}
      \left(\frac{\alpha_{J_\nu}+3}{\alpha_j+3}\right)
      \left(\frac{d_{L}\left(z_j,0\right)}{d_{L}\left(z_j,z\right)}\right)^2
\end{equation}

which is the ratio between the summed photoionization rates of $n$ quasars
at redshifts $z_j$ with a rest frame Lyman limit flux $f_{\nu_\mathrm{LL},j}$,
penetrating the absorber at redshift $z$ and the overall UV background with 
Lyman-limit intensity $J_\nu$.
$d_{L}\left(z_j,0\right)$ is the luminosity distance of QSO $j$, and 
$d_{L}\left(z_j,z\right)$ is its luminosity distance as seen at the absorber; 
the redshift of the quasar as seen at the absorber is $z_j^{\prime}$ \citep{liske00}. 
Thus, $\omega\left(z_\mathrm{em}\right)$ characterizes the expected strength 
of the transverse proximity effect signature at a given QSO redshift along the 
line of sight: 
A highly significant effect would require $\omega\left(z_\mathrm{em}\right) \gg 1$, 
i.e.\ a UV radiation field dominated by the local source.

\begin{table}
\caption{Rest frame Lyman limit fluxes of Q~0302$-$003 and all nearby QSOs. 
A power law $f_{\nu}\propto\nu^{-\alpha}$ is fitted to all quasars with an available spectrum. 
$f_{\nu_\mathrm{LL}}$ is the extrapolated Lyman limit flux in the QSO rest frame and 
$\omega(z_\mathrm{em})$ is the resulting predicted strength of the transverse
proximity effect (see text).}
\label{lylimitfluxes}
\centering
\begin{tabular}{lllll}
\hline\hline \noalign{\smallskip} 
QSO & Abbr.
    & $\alpha$ & $f_{\nu_\mathrm{LL}}$ [$\mu$Jy]
    & $\omega(z_\mathrm{em})$ \\ 
\noalign{\smallskip} \hline \noalign{\smallskip} 
QSO~03022$-$0023 & F & $1.12\pm 0.06$ & $2\pm 4$      & $0.073$ \\
QSO~03027$-$0027 & E & $0.67\pm 0.02$ & $22\pm 17$    & $0.121$ \\
QSO~03027$-$0010 & B & $0.88\pm 0.02$ & $4\pm 4$      & $0.035$ \\
QSO~0302-D113    & D &                & $0.3 \pm 0.3$ & $0.028$  \\ 
QSO~03020$-$0014 & A & $1.42\pm 0.01$ & $10\pm 7$     & $0.253$ \\
Q~0301$-$005     & C & $0.90\pm 0.01$ & $136\pm 89$   & $0.776$ \\
Q~0302$-$003     &$\star$   & $0.50\pm 0.01$ & $222\pm 124$  & $\infty$\\
\noalign{\smallskip} \hline
\end{tabular}
\end{table}

We fixed the UV background at 1 ryd to $J_\nu = 7\times
10^{-22}$~erg~cm$^{-2}$~s$^{-1}$~Hz$^{-1}$~sr$^{-1}$ \citep{scott00} and 
assumed it to be constant over the relevant redshift range $2<z<3.285$ with a
power-law shape $J_\nu\propto \nu^{-\alpha_{J_\nu}}$ and $\alpha_{J_\nu}=1.5$.
The quasar Lyman limit fluxes were estimated from the available spectra by fitting
a power law $f_{\nu}\propto \nu^{-\alpha}$ to the quasar continuum redward of the
Ly$\alpha$ emission line, excluding the emission lines.
Table~\ref{lylimitfluxes} lists the resulting spectral indices and the Lyman limit 
fluxes for the quasars. For QSO~D there was no spectrum available and we
estimated its Lyman limit flux from its $R$ magnitude assuming a power law
index $\alpha=0.5$.

In all cases except for Q~0302$-$003 itself we find peak $\omega$ values
significantly below unity. The highest value is reached by the very luminous
QSO~C, for which \citet{dobrzycki91} proposed that it might cause the large
void at $z\sim 3.17$. While the redshift offset between this void and
QSO~C at $z=3.231$ may still be explained by anisotropic emission, the
small $\omega(z_\mathrm{em})$ value implies that this quasar would
have to be implausibly bright in the direction of the void
\citep{dobrzycki91b}. Note that due to their small redshift difference,
Q~0302$-$003 and QSO~C have similar impact at $z\simeq 3.231$. 
The $\omega(z_\mathrm{em})$ value of QSO~C is therefore the sum 
of the contributions from these two sources. The small 
projected proximity effect zone of QSO~C merges with the onset of the much
larger line of sight proximity effect zone caused by Q~0302$-$003 itself. The
refined redshift of QSO~C implies an even larger offset between quasar and
void; together with the modest value of $\omega$ it follows that the
\citet{dobrzycki91} void is very unlikely related to the ionizing radiation
from QSO~C. This statement will be reinforced below in Sect.~\ref{hardness}.

For all other QSOs the $\omega$ values are even much smaller. Consequently, as
long as one assumes roughly isotropic radiation, these QSOs should leave no
individually detectable traces of a transverse procimity effect. This also
holds for QSO~A. Its matching redshift with a small \ion{H}{i} void could
therefore be just a coincidence, especially when considering the large number 
of similar features in the spectrum. Nevertheless, the combined evidence 
from \ion{H}{i} and \ion{He}{ii} forests suggests that this region does 
indeed receive a substantial amount of hard QSO radiation 
(H00, J03).

\section{Fluctuations in the \ion{He}{ii} Lyman forest} \label{intensity_he2}

Now we turn to the \ion{He}{ii} Ly$\alpha$ absorption observed
with STIS on HST in the redshift range $2.76<z<3.285$ by H00. 
Figure~\ref{he2h1trans} compares the transmission spectra of the
\ion{He}{ii} and \ion{H}{i} Ly$\alpha$ absorption towards Q~0302$-$003
as a function of redshift. The main features are labeled following H00.

There are only three confined regions of significantly non-zero transmission
in the \ion{He}{ii} spectrum. Firstly, the redshift range from $z=3.22$ up to
the emission redshift of Q~0302$-$003 of $z_{\mathrm{em}}=3.285$ displays
strongly enhanced transparency, presumably through a strong line of sight
proximity effect created by a Str\"{o}mgren sphere in an IGM where \ion{He}{ii}
is still largely neutral \citep[][H00]{hogan97}. The luminous QSO~C
is located within the extent of this zone, albeit near its low-redshift end.

Second, there is a 
prominent transmission feature at $3.043 < z < 3.063$ (`void A').
H00 speculated that this was likely to be created by a nearby AGN. 
J03 found the closely coinciding quasar QSO~03020$-$0014 (QSO~A)
and argued that this QSO most likely was responsible for the void, although
they invoked some modest degree of anisotropic radiation in order
to explain an apparent shift between the QSO at $z=3.050\pm 0.003$ and their 
stated peak of void~A at $z=3.056$. From our measurement the peak of void~A is at
redshift $z=3.052$ which is perfectly consistent with the QSO being located 
symmetric to the void.

The third transmission window is a considerably wider region at $z\la 2.87$
(`region B'). The $S/N$ per pixel in the STIS data is low, but the
transmission nearly continuously exceeds the noise. Region B is located at the
low-redshift end of the STIS spectrum and the detector sensitivity strongly
decreases towards the cutoff at $z\sim 2.76$, but it is likely that we see the
onset of a proper \ion{He}{ii} Ly$\alpha$ forest. This interpretation is supported by
the large extent of this region with inherent substructure and a spectral
hardening of the radiation field in region B found by H00.

QSO~B falls into the redshift range of region B, but that region is much too
broad to allow for a close connection between this quasar and the overall
enhancement of transparency. However, we notice that while the \ion{H}{i}
forest near QSO~B actually shows significant absorption 
there is no corresponding depression in the \ion{He}{ii} forest as would be
expected for a roughly constant column density ratio. We tentatively conclude
that this ratio may change significantly near QSO~B, possibly indicating the
radiative influence of the quasar.

\section{The spectral hardness of the UV radiation field} \label{hardness}

\subsection{Diagnostics}

A key parameter to study the radiative influence of a given quasar on the IGM
is the relative hardness of the UV radiation ionizing the
intergalactic matter. There are several established ways to express this
hardness. If the shape of the ionizing continuum is known, the spectral softness
parameter is defined as the ratio between the ionization rates of \ion{H}{i} 
and \ion{He}{ii},
\begin{equation}
S=\frac{\Gamma_\ion{H}{i}}{\Gamma_\ion{He}{ii}}\simeq 4^{\alpha+1}
\end{equation}
with $\alpha$ as the relevant spectral index, $f_\nu \propto \nu^{-\alpha}$.
More indirectly, the hardness can be described by the ratio $\eta$ of the
column densities of \ion{He}{ii} and \ion{H}{i} at given $z$,
\begin{equation}
\eta = \frac{N_{\ion{He}{ii}}}{N_{\ion{H}{i}}}.
\end{equation}
This quantity is theoretically predicted to be $\la 100$ for a hard ionizing
field with mostly quasars acting as ionizing sources, and $\ga 100$ for a soft
radiation field that is dominated by galaxies \citep{haardt96,fardal98};
typically, $S\simeq 2.3\eta$ if \ion{H}{i} and \ion{He}{ii} are highly ionized. 
Unfortunately it is hard to measure $\eta$ directly, because of the limited 
spectral resolution of the FUV spectrographs aboard HST. 
Only for the two lines of sight towards HE~2347$-$4342 and HS~1700$+$6416, 
FUSE spectra of sufficient resolution have allowed direct measurements of 
\ion{He}{ii} column densities and $\eta$ \citep{kriss01,zheng04,shull04,reimers04}, 
revealing fluctuations of $\eta$ on very small scales of $\Delta z\approx10^{-3}$, 
indicating the presence of hard and soft photoionizing sources whose radiation is 
filtered through the cosmic web.

At lower resolution a technique has been applied that allows one to at least
roughly estimate $\eta$. Using accurately measured \ion{H}{i} column densities
from optical spectroscopy, one assumes a value for $\eta$, generates the
ensuing \ion{He}{ii} spectrum, degrades it to the actual resolution of the UV
spectrum and compares the simulation with the data
\citep{reimers97,hogan97,anderson99,heap00,smette02}. These studies showed
consistently and in agreement with the higher resolution FUSE results that
$\eta$ significantly fluctuates over the entire redshift range covered by a
given line of sight. On this basis H00 predicted that void~A in Q~0302$-$003
had to be created by a QSO because only a small value of $\eta\sim 50$ 
could reproduce the STIS spectrum in this region, whereas outside of it 
$\eta$ was much larger. Subsequently J03 discovered QSO~A almost exactly at
the predicted redshift.

An alternative to the above procedure is given by evaluating the ratio $R$ of 
the effective optical depths of \ion{He}{ii} and \ion{H}{i},
\begin{equation}
R \equiv\frac{\tau_{\mathrm{eff},\ion{He}{ii}}}{\tau_{\mathrm{eff},\ion{H}{i}}},
\end{equation}
which has the same overall characteristics as $\eta$, but is resolution-independent. 
High (small) values of $R$ will be obtained if the \ion{He}{ii} absorption is 
high (small) compared to \ion{H}{i}, indicating a soft (hard) photoionizing field.

We now want to investigate fluctuations of $\eta$ in other parts of the 
Q~0302$-$003 spectrum, notably in region~B. Our approach is therefore
reverse to that of H00 and J03, in that we have now specific redshifts
given by quasars, and we are interested in  the behaviour of
$\eta$ at these redshifts. However, we first need to consider the
question of how much these results depend on unresolved weak Lyman forest lines.

\subsection{Unresolved weak \ion{H}{i} forest lines}

For given \ion{H}{i} column density, the \ion{He}{ii} absorption is typically
much stronger than the corresponding \ion{H}{i} absorption ($\eta\gg 1$).
\ion{He}{ii} can therefore be traced in absorbers with $\log{N_\ion{H}{i}} \la
12$, where the more diffuse component of the IGM sets in -- effectively a
shallow Gunn-Peterson trough \citep[e.g.][]{songaila95,fardal98}. This IGM
component is very hard to account for in the \ion{H}{i} forest because its
detection relies on small shallow excursions of the data from the assumed
quasar continuum that can easily get lost in the fitting process.
Almost every quasar continuum in high-resolution data is defined
locally, so one typically does not account for this low-density IGM
component. This poses a fundamental problem when trying to estimate the
spectral hardness of the UV radiation field by comparing the absorption in
\ion{H}{i} and \ion{He}{ii}. The derived values for $\eta$ and $R$ will be
systematically overestimated when the true \ion{H}{i} absorption is
underestimated.

The best approach to account for the additional undetected \ion{H}{i}
absorption is by resolving the \ion{He}{ii} absorption into discrete lines and
using the \ion{He}{ii} forest to predict the low end of the \ion{H}{i} column
density distribution \citep{kriss01,zheng04}. This option is available only
for two quasar lines of sight with FUSE data.

Other possibilities are to use analytic approximations to calculate the mean 
Ly$\alpha$ forest absorption arising from these lines \citep{moller90,zuo93,madau95} 
or to simulate their impact with empirically established distribution laws.
Noting that the line lists of \citet{hu95} and \citet{kim02} become grossly 
incomplete below $\log{N_\ion{H}{i}} \simeq 12$, we estimated the 
contribution of $9\le \log{N_\ion{H}{i}}\le 12$ systems to the total absorption.
This yields a lower limit on $\tau_\mathrm{GP}$ from completely undetected absorption 
lines. We assumed a power law-shape column density distribution
function with slope $\beta = 1.5$ and used the empirical redshift
distribution parameterization from \citet{kim02}. From both simulation and analytic 
approximation we obtained an effective optical depth of $\tau_\mathrm{eff}\simeq 0.02$ 
for this column density range.

A suitable correction has to be higher than this lower limit due to the partial 
incompleteness of the line lists at $\log{N_\ion{H}{i}}\lesssim 13$. H00 generated 
weak forest lines with $9\le \log{N_\ion{H}{i}}\le 13$ by extrapolating the 
distribution functions from \citet{kim97}, resulting in an estimate of 
$\tau_\mathrm{GP}\simeq 0.06$. Our best-guess value is somewhat lower than this; 
based on exploring various parameterizations of the distribution functions for the 
range $12\le \log{N_\ion{H}{i}}\le 13$, we adopted an overall $\tau_\mathrm{GP} \simeq 0.04$.

We applied this correction over the whole redshift range of interest by multipying our 
FORS2 transmission spectrum by a factor $\left<T_{\ion{H}{i},\mathrm{GP}}\right>= 0.96$. 
The main source of uncertainty of both used approaches lies certainly in the 
extrapolation of the column density distribution to below the validated range. 
However, this is not a major problem for our study because we are primarily interested 
in spatial fluctuations of $R$ and $\eta$ rather than absolute values.

\begin{figure}
\centering 
\includegraphics[width=70 mm,height=51 mm]{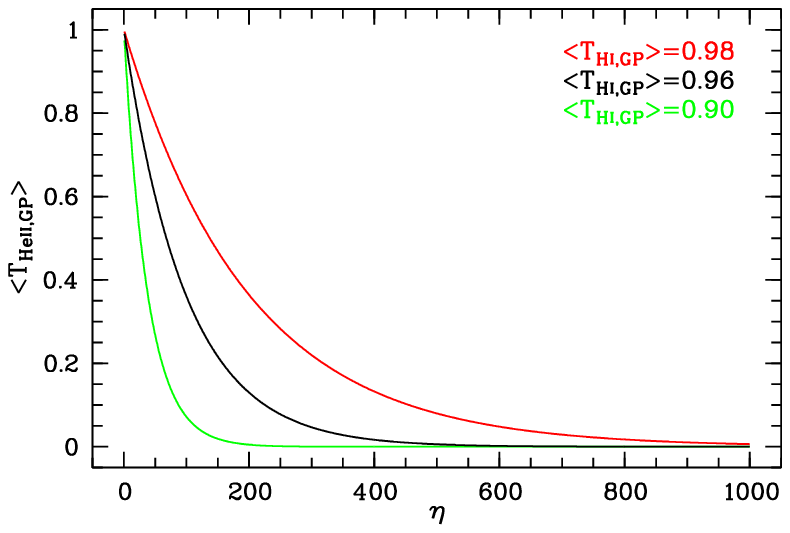}
\caption{\label{transgp_eta}$\left<T_{\ion{He}{ii},\mathrm{GP}}\right>$
vs. $\eta$ for different $\left<T_{\ion{H}{i},\mathrm{GP}}\right>$. In the
range of physical $\eta$ values, the \ion{He}{ii} transmission correction is
strongly dependent on $\left<T_{\ion{H}{i},\mathrm{GP}}\right>$. \emph{[See
the online edition of the Journal for a colour version of this figure.]}}
\end{figure}

The additional absorption in \ion{He}{ii} Ly$\alpha$ that arises from the
cloud population not observed in \ion{H}{i} can be easily calculated. These
absorbers are optically thin both in \ion{H}{i} and \ion{He}{ii}, so that
\begin{equation}
\left<T_{\ion{He}{ii},\mathrm{GP}}\right>=\left<T_{\ion{H}{i},\mathrm{GP}}\right>^{\eta/4}.
\end{equation}
Applying the \ion{H}{i} transmission correction factor gives a global
\ion{He}{ii} transmission correction as a function of $\eta$. Simulated
\ion{He}{ii} spectra created from incomplete \ion{H}{i} line lists have to be
multiplied by $\left<T_{\ion{He}{ii},\mathrm{GP}}\right>$. Since the depth of
the shallow \ion{H}{i} Gunn-Peterson trough is not exactly known, a degeneracy
in the hardness parameter $\eta$ arises due to the strong dependence of the
$\eta$ value fitting the data on the assumed
$\left<T_{\ion{H}{i},\mathrm{GP}}\right>$. From Fig.~\ref{transgp_eta} we find
that all simulated \ion{He}{ii} spectra except those with very low
($\eta\simeq 1$) or very high assumed $\eta$ depend strongly on the value of
$\left<T_{\ion{H}{i},\mathrm{GP}}\right>$. Any study based on low-resolution
\ion{He}{ii} spectra will therefore be plagued by the ambiguity that different
$\eta$ values will fit the data, depending on the adopted correction for
low-column density material. The best-fitting $\eta$ will overestimate
(underestimate) the true value if $\left<T_{\ion{H}{i},\mathrm{GP}}\right>$ is
set too high (low). As a consequence of this degeneracy, the spectral hardness
values derived from low-resolution \ion{He}{ii} data have an uncertain
absolute scale.  However, any indications for fluctuations in the fitting
$\eta$ values along the line of sight due to a changing hardness of the
ionizing field will remain valid, unless the shallow Gunn-Peterson trough
correction itself fluctuates with a similar rate.

\subsection{The fluctuating optical depth ratio} \label{hardness_r}

\begin{figure*}
\sidecaption
\includegraphics[width=12cm]{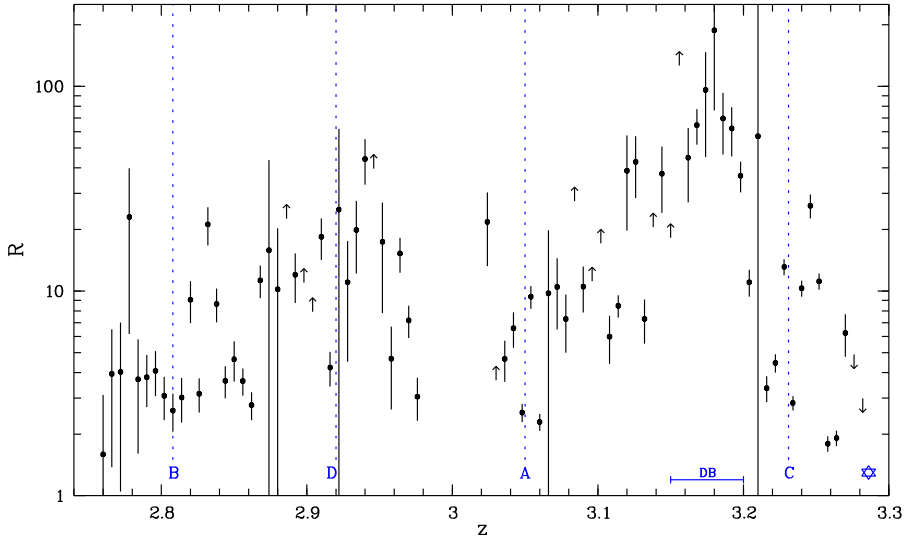}
\caption{\label{forestratio}
Ratio of effective optical depths $R$ vs. redshift $z$. Arrows indicate upper 
and lower limits. Quasar and void locations are marked as in Fig.~\ref{q0302forest}. 
Note that $R$ reaches small values near every quasar along the line of sight. 
There is a distinctive local minimum in $R$ at the redshift of QSO~B. 
\emph{[See the online edition of the Journal for a colour version of this figure.]}}
\end{figure*}

In order to apply the above defined hardness indicator $R$,
we binned the STIS spectrum and the FORS2 spectrum of Q~0302$-$003 into
aligned redshift bins of common size. For the STIS spectrum we adopted the
wavelength resolution of $\Delta\lambda\simeq 1.8$~\AA\ as estimated by
H00 and binned both transmission spectra into $\Delta z=0.006$ bins
starting at $z=2.76$ up to the emission redshift of Q~0302$-$003 at
$z=3.285$. This rebinning implies that one has to deal with original flux
bins that only partly overlap with the new bins. We adopted the method from
\citet{telfer02}, who weighted each original flux by the extent of the overlap
with the new bin. The errors were computed accordingly. 

The optical depth ratio $R=\ln{\left<T_{\ion{He}{ii}}\right>}/\ln{\left<T_{\ion{H}{i}}\right>}$
would obtain unphysical values if $\left<T\right>\le0$ or $\left<T\right>\ge1$. 
In our case only the \ion{He}{ii} spectrum is affected. Two $\left<T_{\ion{He}{ii}}\right>\ge1$ 
bins near Q~0302$-$003 are contaminated by its blue \ion{He}{ii} emission line wing. 
Due to their location in the proximity effect zone, we estimate 
$\left<T_{\ion{He}{ii}}\right>\gtrsim 0.9$, corresponding to upper limits on $R$. 
Values with $\left<T_{\ion{He}{ii}}\right>\le0$ have been replaced by their errors, 
yielding upper limits on $\left<T_{\ion{He}{ii}}\right>$ and lower limits on $R$.
The region $2.98<z<3.02$ is excluded because of the contamination from the 
geocoronal Ly$\alpha$ emission line in the STIS spectrum. Error bars for $R$ were 
computed by propagating the transmission errors.

The result for all created redshift bins is shown in
Fig.~\ref{forestratio}. One clearly sees a large variation in $R$, ranging
from values of a few to $\ga 50$, indicating substantial spectral fluctuations
in the UV radiation field. Most interestingly, $R$ reaches a local minimum
near every quasar along the line of sight, as expected for hard UV
sources. Specifically, $R\simeq 3$ near all known quasars in the field except
maybe QSO~D, for which we have only a single bin at $z=2.916$ with a small
value of $R\simeq 4$. Notice that $R\simeq 3$ also close to
Q~0302$-$003 itself, and again so at the redshift of QSO~C.

In order to estimate the statistical significance to find a local $R\lesssim 3$ 
minimum near the foreground quasars we performed Monte Carlo simulations.
By randomly distributing quasars over the considered redshift range we estimate 
from 10000 simulations a probability of $\sim 9.3$\% that a single quasar falls 
into a redshift bin that corresponds to a local $R$ minimum at $\log(R)<0.5$. 
Since three foreground quasars (QSO~A, B and C) fall in such a minimum, the 
probability to find this constellation by chance is $<0.1$\%.

Figure~\ref{forestratio} shows also that the radiation field in the 
\citet{dobrzycki91} \ion{H}{i} void appears to be very soft, as already 
observed by H00. The errors are substantial, however, because
throughout the void there is still a Gunn-Peterson trough seen in
\ion{He}{ii}, with only very weak absorption in \ion{H}{i}.
There is a clear trend of higher $R$ values when approaching the approximate
center of the void at $z\simeq 3.175$, reaching up to $R\sim 200$. 

Finally we note that in Fig.~\ref{forestratio} there are three narrow redshift
regions at $z=2.778$, $z=2.832$ and $z=2.940$ where the local radiation field
seems to be quite soft with $R \ga 20$.  From Fig.~\ref{he2h1trans} we see
that these high $R$ values correspond to small \ion{H}{i} voids. Such a
correlation has been observed already in the high-resolution studies with FUSE
towards HE~2347$-$4342 \citep{kriss01,shull04} and may indicate the
contribution of star-forming galaxies to the UV background or quasar radiation
that has been filtered and softened by radiative transfer through the cosmic
web. The HST data for Q~0302$-$003 are insufficient to assess this in any
detail.

\subsection{The fluctuating column density ratio} \label{hardness_eta}

We now turn to the more indirect approach of estimating $\eta$ through
comparing the predicted with the observed \ion{He}{ii} absorption. 
We used the \ion{H}{i} line lists of \citet{hu95} and \citet{kim02}, 
which jointly cover the redshift range accessible with STIS in \ion{He}{ii} 
Ly$\alpha$. The \citet{hu95} list covers the range $2.627\le z\le 3.110$, 
where the QSOs B, D, and A are located, whereas the list from
\citet{kim02} extends between $2.957\le z\le 3.235$, including 
QSOs A and C, the \ion{H}{i} void by \citet{dobrzycki91}. Only a small
part of the line of sight proximity effect zone of Q~0302$-$003 itself
is covered.

We simulated the Voigt profiles of the \ion{He}{ii} Ly$\alpha$ transition by
assuming pure non-thermal broadening of the lines 
($b_\ion{He}{ii}=b_\ion{H}{i}$) and a constant value for $\eta$ along the line
of sight that was used to convert the observed $N_{\ion{H}{i}}$ into 
$N_{\ion{He}{ii}}$. Non-thermal broadening found in simulations can be either 
due to turbulent motions of the gas or the differential Hubble flow, with the 
latter being dominant for the low-column density forest 
\citep{zhang95,zhang98,hernquist96,weinberg97,bryan99}. 
Additionally \citet{zheng04} presented observational evidence for 
$b_\ion{He}{ii}\simeq b_\ion{H}{i}$. We then applied the \ion{He}{ii} 
transmission correction factor for undetected material in \ion{H}{i}
$\left<T_{\ion{He}{ii},\mathrm{GP}}\right>\simeq 0.96^{\eta/4}$. We performed
the simulations for a range of $\eta$ values from 15 to 3000. Finally, we
degraded the resolution of the \ion{He}{ii} transmission spectrum to the
actual STIS resolution by convolving it with a Gaussian line spread function
of $1.8$~\AA\ FWHM and rebinning the spectrum to the STIS pixel size of
$0.6$~\AA.

\begin{figure*}
\centering \includegraphics[scale=0.95]{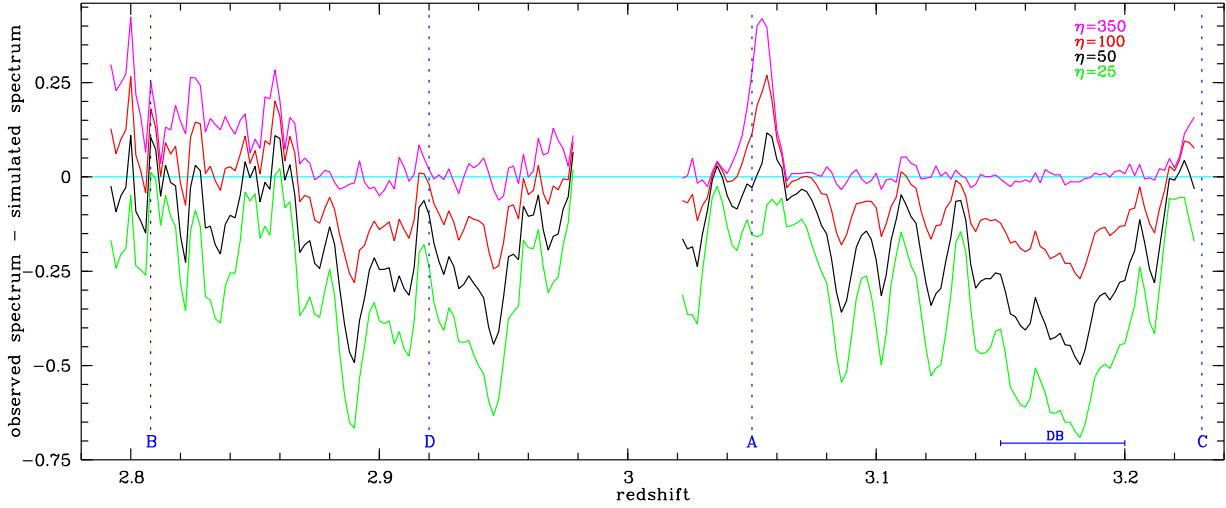}
\caption{\label{diff_eta}
Observed vs.\ predicted \ion{He}{ii} Ly$\alpha$ absorption, in differential
representation (observed minus predicted). Different values for $\eta$ were used 
for the simulations (see text). The part of the spectrum with $z\le 2.98$ was simulated 
using the \ion{H}{i} line list by \citet{hu95}, whereas the \ion{H}{i} line list by 
\citet{kim02} was used at $z\ge 3.02$. The spectral region contaminated by geocoronal 
Ly$\alpha$ emission is not shown. Quasar and void locations are marked as in Fig.~\ref{q0302forest}. 
Positive (negative) deviations from zero indicate that $\eta$ has to be smaller (higher) than the 
assumed $\eta$ of the curve. 
\emph{[See the online edition of the Journal for a colour version of this figure.]}}
\end{figure*}

The results for four representative values of $\eta$ are shown in
Fig.~\ref{diff_eta} as the difference between the observed and the simulated
spectrum, where positive (negative) deviations from zero indicate that $\eta$
has to be smaller (higher) than the assumed $\eta$ of the curve. One can
clearly see that different values of $\eta$ match to different redshift
regions, so the hardness of the ionizing radiation field has to fluctuate
along the line of sight. Remarkably, the spectral regions around most of 
the quasars are best reproduced by low $\eta$ values, which are typical of 
quasars \citep{haardt96,fardal98}.

The onset of the line-of-sight proximity effect region of Q~0302$-$003 is
reproduced with $\eta\sim 50$. H00 analyzed the proximity effect
zone of Q~0302$-$003 in detail and found that $\eta$ strongly decreases when
approaching Q~0302$-$003. Thus, in the proximity effect zone of Q~0302$-$003
hard radiation is present that is softened with increasing distance to the
quasar. 

In the nearby \ion{H}{i} void found by \citet{dobrzycki91}, 
a physical underdensity or a very soft radiation field with 
$\eta\ga 350$ is required, as already noted by H00. 
Such high $\eta$ values are inconsistent with quasars as main ionizing 
sources. Furthermore, neither the survey by J03 nor our survey did reveal 
previously unknown quasars close to the void. 
As discussed in Sect.~\ref{trprox_h1} above, QSO~C is presumably not capable
of creating the void. There is also no large 
overdensity of LBGs observed in the redshift range of the void
that might photoionize the \ion{H}{i} \citep{adelberger03}.
Most likely, this void is a large-scale structure feature 
and not one created by local photoionizing sources.

J03 associated QSO~A with void A mainly by showing
that such an association has a very small probability of chance occurrence, 
and that the luminosity of QSO~A is high enough to account for the
void. Figure \ref{diff_eta} shows clearly that most parts
of void A can be fitted with $25 \la \eta\la 50$, which clearly
favours a transverse proximity effect from a quasar.

\begin{figure}
\centering \includegraphics[scale=1]{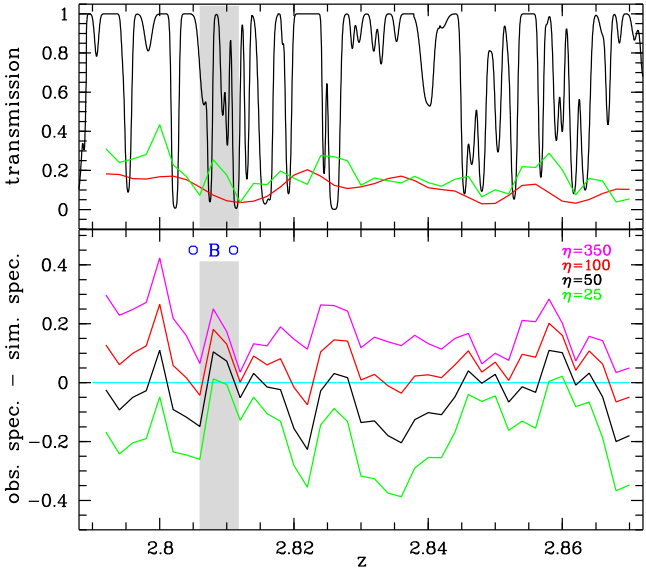}
\caption{\label{diff_eta3}
Close-up of region B ($z<2.87$). The upper panel shows the \ion{H}{i} forest 
spectrum of Q~0302$-$003 reconstructed from the line list by \citet{hu95} (black), 
the \ion{He}{ii} transmission (green) and the simulated STIS spectrum for 
$\eta=100$ (red). The lower panel shows the fluctuating spectral hardness 
in this region (close-up of Fig.~\ref{diff_eta}). The blue letter marks the 
foreground quasar QSO~03027$-$0010 (QSO~B) at $z=2.808$. Blue circles indicate 
the two nearby LBGs found by \citet{steidel03}. In the grey-shaded redshift range 
$2.806\le z\le 2.812$, $\eta=$25--50 provides the best fit. 
\emph{[See the online edition of the Journal for a colour version of this figure.]}}
\end{figure}

Region B is quite crowded, and in Fig.~\ref{diff_eta3} we present an enlarged
view of Fig.~\ref{diff_eta} for this redshift range. 
The best-fit value of $\eta$ fluctuates between
$\eta \sim 100$ and $\sim 25$ on scales comparable to the spectral resolution 
of the STIS data. This overall hardening of the metagalactic UV radiation
field compared to the soft radiation field in the Gunn-Peterson trough 
can probably be attributed to the progressing \ion{He}{ii} reionization
and the onset of a proper \ion{He}{ii} Lyman forest \citep{zheng04}.

Notably, $\eta\simeq 25$ is reached only in two points,
one of which precisely concides with the redshift where our newly discovered 
QSO~B is located. This match is very unlikely to be a chance occurence; it is
much more plausible to assume that hard radiation from QSO~B intercepts the
line of sight towards Q~0302$-$003 at this point.

\citet{steidel03} found two LBGs at angular distances of $\vartheta=3\farcm 04$
and $\vartheta=4\farcm 33$ from Q~0302$-$003 near the redshift of QSO~B, at $z=2.805$ 
and $z=2.811$ (also indicated in Fig.~\ref{diff_eta3}).
While the \ion{H}{i} Lyman continuum escape fraction of LBGs is still debated
\citep{steidel01,giallongo02,fernandez-soto03,shapley03}, the number of
\ion{He}{ii} ionizing photons produced is at most a few per cent 
of the number of photons exceeding the \ion{H}{i} Lyman edge 
according to current evolutionary synthesis models of starburst galaxies
\citep{leitherer99,smith02,schaerer03}. 
We therefore rule out the possibility that these two LBGs supply the hard 
photons required for the inferred $\eta$ minimum.

The situation is less clear for the weak QSO~D also discovered by
\citet{steidel03}. At the redshift of this QSO, the overall appearance
of the \ion{He}{ii} forest is that of a completely opaque Gunn-Peterson
trough, probably reflecting a still partly neutral \ion{He}{ii} component of
the IGM. This region is well described with $\eta \sim 350$
(Fig.~\ref{diff_eta}). However, we see also that Fig.~\ref{diff_eta}
suggests a drop down to $\eta\sim 100$ again at exactly the redshift
of a QSO near the line of sight. This corresponds to a small
spike of excess flux in the Gunn-Peterson trough 
(see Fig.~\ref{he2h1trans}) which is formally still consistent 
with the noise level. While we therefore cannot really claim to
have detected another case of the transverse proximity effect,
we cannot exclude it either, and the coincidence is certainly
suggestive.

\section{Discussion} \label{disc}

Within the redshift range covered by the STIS data on the \ion{He}{ii} forest
of Q~0302$-$003, five QSOs are known in or along the line of sight. Only one
of these, the central quasar Q~0302$-$003 itself, shows unambiguous evidence
for a proximity effect in \ion{H}{i}. There is a possible small void near
QSO~A, but this could just as well be due to the overall distribution of
\ion{H}{i} material along the line of sight. At the redshifts of all three 
other QSOs, the transmission of the \ion{H}{i} Lyman forest in the line of sight 
towards Q~0302$-$003 seems to be rather below average. Our estimated
UV luminosities of these QSOs confirm that a classical transverse proximity
effect is not really expected.

Considering the STIS UV spectra of the \ion{He}{ii} Lyman forest region
instead, this picture does not change dramatically. Only in one case, at the
redshift of QSO~A, we see a clear `\ion{He}{ii} void', which in fact led to the
discovery of the QSO (H00, J03).  For QSO~B and maybe also for QSO~D we
find nonzero \ion{He}{ii} transmissions, but these certainly do not qualify
as voids in their own right. A similar case holds for QSO~C (see
Fig.~\ref{he2h1trans}), where however a confusion with the line of sight
proximity effect zone of Q~0302$-$003 complicates the situation.

Only by combining \ion{H}{i} and \ion{He}{ii} transmission properties,
the radiative imprints of the QSOs along the line of sight become clearly
separated from large scale fluctuations of the distribution of matter.
This is particularly prominent for region~B and QSO~B, where the \ion{H}{i} 
forest of Q~0302$-$003 shows in fact a complex of enhanced \ion{H}{i}
absorption, possibly even associated with the QSO in form of a large scale
structure filament (Fig.~\ref{diff_eta3}). Eleven absorbers with 
$\log{N_\ion{H}{i}}>13$ reside in the redshift range $2.80<z<2.82$, 
four of them between the projected positions of the LBGs at $z=2.805$ and $z=2.811$.
Assuming an average spectral hardness in this redshift
regime ($\eta\sim 100$), one expects strong corresponding \ion{He}{ii} absorption
arising from these absorbers even at the low resolution of STIS (Fig.~\ref{diff_eta3}).
Instead, we see that the \ion{He}{ii} forest remains transparent at this redshift, 
however with nothing that could be interpreted as a `void'. Both spectral 
hardness estimators $R$ and the best-fit $\eta$, on the other hand, show a clearly 
pronounced extremum at precisely the correct redshift, in excellent agreement with
the easier to interpret cases of QSO~A and Q~0302$-$003. Thus, the most likely 
reasons for the different appearance of the \ion{He}{ii} transmission portions 
near QSO~A (\ion{He}{ii} void) and QSO~B (no void) in spite of a similar inferred 
hardness are different intrinsic densities of the overionized regions 
(small \ion{H}{i} void vs. filament) combined with the progressing \ion{He}{ii} 
reionization with lower fluctuations of the \ion{He}{ii} transmission and the 
spectral hardness at lower redshift. 
The same effect, though much less significant, can be seen for QSO~D.

Even for QSO~C there might be a similar imprint onto the Q~0302$-$003 line of
sight. Given the UV luminosity of Q~0302$-$003, the zone of overionized
\ion{He}{ii} in the line of sight is surprisingly large, especially if the
surrounding IGM was still mostly neutral in \ion{He}{ii}.  A possible
additional contributor of hard ionizing photons could be the very luminous
QSO~C, intercepting the low-$z$ end of the proximity effect region of
Q~0302$-$003.  Figures \ref{forestratio} and \ref{diff_eta} provide evidence
for a very hard radiation field at just the redshift of QSO~C which then,
towards Q~0302$-$003, becomes somewhat softer again before finally hardening
in the inner proximity effect zone. We suggest that the
\ion{He}{ii} void at $z>3.22$ is really generated by the combined actions of a
transverse proximity effect of QSO~C and a line of sight effect of
Q~0302$-$003. 

We conclude that there is strong evidence for a \emph{transverse proximity 
effect in spectral hardness} occuring in several incidences along the line 
of sight towards Q~0302$-$003. While individual $\eta$ fluctuations might
still be regarded as coincidental, and possibly due to the crude diagnostics
used, local minima of $R$ and $\eta$ are in fact found at exactly the 
redshifts of all five relevant QSOs, indicating a UV radiation field 
significantly enhanced due to a local hard source. Notice that because of the 
low overall UV background at 4~ryd, the proximity effect zone in spectral
hardness is considerably larger than that of the conventional \ion{H}{i}
proximity effect. For QSO~B, it reaches over $\sim 5$~Mpc, and for the
luminous QSO~C it may even bridge more than 10~Mpc. 

Besides the continuum placement and the background subtraction, the main 
source of uncertainty in our assessment of $R$ and $\eta$ is the 
correction for weak unresolved \ion{H}{i} lines, the `shallow
\ion{H}{i} Gunn-Peterson trough' created by the quasi-diffuse component in 
the IGM. Small changes in the adopted correction lead to large changes 
in the best-fit $\eta$, as $\eta$ is very sensitive to 
small \ion{H}{i} column densities. The same is true for the optical depth ratio 
$R$, which is very sensitive at high \ion{H}{i} transmission values. 
This degeneracy affects the inferred $R$ and $\eta$ values as such,
but the observed \emph{fluctuations} and their correlation with the
redshifts of known quasars are largely unaffected. Furthermore, the range of
our inferred $\eta$ values ($25 \la \eta \la 350$) is generally believed to
be typical for the conditions in the IGM at $z \sim 3$. 

A very interesting, though highly speculative observation concerns the
relation between the strength of the `spectral hardness proximity effect' and
the UV spectral slope of the QSO in question. QSO~A is about 1~mag brighter in
$V$ than QSO~B, and it is also somewhat closer to the central line of sight.
On the other hand, the inferred $\eta$ is even slightly lower at the redshift 
of QSO~B than in void~A. Adopting the continuum slopes given in 
Table~\ref{lylimitfluxes} and boldly extrapolating towards 4~ryd, 
we predict values of the spectral softness of $S\simeq 29$ for QSO~A 
and $S\simeq 10$ for QSO~B; thus, the latter QSO is expected to be even
\emph{brighter} at the \ion{He}{ii} ionizing limit than the former. 
The higher $\eta$ values are therefore inferred near QSO~A
which has the softer spectrum, whereas lower $\eta$ values fit 
the data near QSO~B which has the harder spectrum.

\section{A lower limit on the quasar lifetime} \label{lifetime}

Both incarnations of the proximity effect allow a lower limit on the quasar
lifetime $t_{\mathrm{q}}$, since photons from a quasar need a certain time to
propagate to the point of their absorption in the IGM. The observation of the
line-of-sight proximity effect in the Ly$\alpha$ \emph{forest} yields only
lower limits based on the typical equilibrium timescale of Ly$\alpha$
absorbers of $\simeq 10^4$~yr \citep[e.g.][]{bajtlik88}, mainly because the
line-of-sight proximity effect observed in a line forest is a statistical
effect. However, in a mostly neutral medium, there exists the sharp observable
boundary that defines the radius of the Str\"{o}mgren sphere around the
quasar. So far, this has been observed in \ion{H}{i} Ly$\alpha$ towards
\object{SDSS~J1030$+$0524} at $z=6.28$ \citep{pentericci02} and in
\ion{He}{ii} Ly$\alpha$ towards Q~0302$-$003 \citep{hogan97} and
PKS~1935$-$692 \citep{anderson99}. Simple modelling of the zone inside the
Str\"{o}mgren sphere yields $t_{\mathrm{q}}\gtrsim 13$~Myr
\citep{pentericci02} and $t_{\mathrm{q}}\gtrsim 40$~Myr
\citep{hogan97}.

In the case of the transverse proximity effect one cannot measure the extent
of the proximity effect zone, neither in a line forest nor in a Gunn-Peterson
trough, since the proximity effect zone is intersected by another line of
sight by chance. The emission epoch of the ionizing photons of the source in
the direction of the background line of sight cannot be observed, since these
photons have to be emitted prior to our observations. So the transverse proper
distance $d_\perp$ between the two lines of sight at the emission redshift of
the foreground quasar has to serve as an approximation for the expanding
distance between the source and the emitted photons. Due to the expansion of
the universe between the emission and the absorption of the photons, the
inferred light travel time $\Delta t^\prime=d_\perp/c$ is larger than the
actual light travel time $\Delta t$, but usually the correction for
expansion is very small.

We have argued in the previous section that the QSOs~A--D show at least an 
indication for the transverse proximity effect in the line of sight towards 
Q~0302$-$003. Assuming a correspondence between the known foreground quasars 
and \ion{He}{iii} zones around them having low $\eta$, we can infer their 
minimum total quasar lifetimes listed in Table~\ref{qsolifetimes}.
These are all consistent with the estimates based on the proximity effect 
\citep[][J03]{srianand97,hogan97,anderson99,pentericci02} and theoretical
arguments (\citealt{martini04} and references therein). Due to the high 
sensitivity of the spectral hardness we are able to constrain the quasar 
lifetime for a sample of quasars forming a close group on the sky, yielding 
minimum quasar lifetimes in the range
$\Delta t^\prime\gtrsim 10 \mbox{--}30\, \mathrm{Myr}$.

\begin{table}
\caption{Minimum lifetimes $\Delta t^\prime=d_\perp/c$ for all foreground 
quasars of Q~0302$-$003 with an indication of the transverse proximity effect.}
\label{qsolifetimes}
\centering
\begin{tabular}{lllll}
\hline\hline \noalign{\smallskip} 
QSO & Abbr.
    & $z$ & $d_\perp$ [Mpc]
    & $\Delta t^\prime$ [Myr] \\ 
\noalign{\smallskip} \hline \noalign{\smallskip} 
QSO~03027$-$0010 & B & $2.808$ & $5.29$  & $17$ \\
QSO~0302-D113    & D & $2.920$ & $2.26$  & $7$  \\ 
QSO~03020$-$0014 & A & $3.050$ & $2.97$  & $9$ \\
Q~0301$-$005     & C & $3.231$ & $10.34$ & $34$ \\
\noalign{\smallskip} \hline
\end{tabular}
\end{table}

\section{Conclusions} \label{conclusions}

The transverse proximity effect in quasar spectra has so far mostly been
associated with the notion of voids in the \ion{H}{i} Lyman forest,
created by the overionized zones around quasars near the line of sight.
It has always been clear, however, that true voids due to large scale
structure and apparent voids due to the proximity effect would be essentially
indistinguishable on an individual basis. This problem holds also for the
\ion{He}{ii} Lyman forest that has now become accessible for a small number of
high-redshift QSOs.

We argue in this study that spectral hardness indicators provide an efficient
way to discriminate between large scale structure and proximity effect.  The
line of sight proximity effect towards Q~0302$-$003 and the transverse
proximity effect due to QSO~03020$-$0014 (dubbed `QSO~A' in this paper) were
already discussed by \citet{heap00} and \citet{jakobsen03} and are, 
since the discovery of QSO~A, clear-cut cases where both the optical depth ratio 
and the hardness parameter $\eta$ show pronounced minima close to the quasars. 
We investigated the \ion{H}{i} and \ion{He}{ii} absorption properties at redshifts 
near our newly discovered QSO~03027$-$0010 (`QSO~B'), and also near the other known QSOs
close to the line of sight.  In each single case we find evidence for a 
link between spectral hardness and the presence of a nearby quasar.

Thus, the relative UV hardness is a sensitive \emph{physical} quantity to
search for individual sources of the metagalactic UV radiation field beyond
the simple detection of associations between quasars and voids. Quasars and
voids may be unrelated, whereas the spectral hardness is related to the
spectral energy distributions of the ionizing sources. In particular, a void
will not occur if the quasar overionizes an intrinsically overdense
region. The spectral hardness may break this density degeneracy that affects
studies of the proximity effect \citep{loeb95,schirber04}. At least for
QSO~B, this case of an overdense region is apparently applicable. 

The excellent match between the measured QSO redshifts and the inferred minima 
of $\eta(z)$ imply that there is essentially no need to invoke strongly
anisotropic radiation for any of the quasars in question. This is in stark
contrast to the situation for the \ion{H}{i} proximity effect where anisotropy
or even beaming has repeatedly been claimed 
\citep{dobrzycki91,crotts89,moller92,schirber04}. It is likely that 
true large scale structure in the Lyman forest has contributed to mask 
the intrinsic transverse proximity effect in several of these cases.

The redshift range of $2.7 \la z < 3.2$ studied in the line of sight towards 
Q~0302$-$003 is of very high interest in the context of studying the putative
reionization epoch of \ion{He}{ii}. We have shown that one can determine the 
sources that ionize this line of sight, even at low resolution. Near the
low-redshift end of the covered spectral range, there are clear indications
for a transition from \ion{He}{iii} bubbles in the IGM towards a more or less
fully ionized \ion{He}{iii} IGM where also the overall UV background is
dominated by QSO radiation. Our estimates of $\eta$ in this redshift range
are fully compatible with this concept. Even then, local sources of very hard
radiation such as QSO~B can significantly modify the UV background over
distances of several Mpc. It would be highly desirable to be able to study
these processes at higher spectral resolution. Unfortunately, Q~0302$-$003
is too faint for observations with the \emph{Far Ultraviolet Spectroscopic
Explorer}, and the only perspective for significantly improved data lies with
the Cosmic Origins Spectrograph to be installed on HST.

\begin{acknowledgements}

We thank the staff of the ESO observatories La Silla and Paranal for their
professional assistance in obtaining the optical data discussed in this
paper. We are grateful to Aldo Dall'Aglio for providing his code to simulate 
absorption line spectra from a given line list. GW acknowledges support by a
HWP grant from the state of Brandenburg, Germany.\\
Funding for the creation and distribution of the SDSS Archive has been
provided by the Alfred P. Sloan Foundation, the Participating Institutions,
the National Aeronautics and Space Administration, the National Science
Foundation, the U.S. Department of Energy, the Japanese Monbukagakusho, and
the Max Planck Society. The SDSS Web site is http://www.sdss.org/.
The SDSS is managed by the Astrophysical Research Consortium (ARC) for the
Participating Institutions. The Participating Institutions are The University
of Chicago, Fermilab, the Institute for Advanced Study, the Japan
Participation Group, The Johns Hopkins University, the Korean Scientist Group,
Los Alamos National Laboratory, the Max-Planck-Institute for Astronomy (MPIA),
the Max-Planck-Institute for Astrophysics (MPA), New Mexico State University,
University of Pittsburgh, University of Portsmouth, Princeton University, the
United States Naval Observatory, and the University of Washington.

\end{acknowledgements}

\bibliography{phdthesis.bib}
\end{document}